
%
%

%
%
%
%

\def\Serif{cmr}
\def\SerifBold{cmbx}
\def\SerifItalics{cmti}
\def\SerifSlanted{cmsl}
\def\SerifBoldItalics{cmbxti}
\def\SansSerif{cmss}
\def\SansSerifBold{cmssbx}
\def\SansSerifItalics{cmssi}
\def\SansSerifSlanted{cmssi}
\def\Math{cmmi}
\def\Symbols{cmsy}
\def\MathBold{cmmib}
\def\MoreSymbols{cmex}
\def\Typewriter{cmtt}
\def\Gothic{eufm}
\def\Double{msbm}
\def\Relazioni{msam}

= 			\Serif10 			at 5pt
= 		\SerifBold10 		at 5pt
= 	\SerifItalics10 	at 5pt
=		\SerifSlanted10 	at 5pt
=	\SerifBoldItalics10	at 5pt
= 		\SansSerif10 		at 5pt
=	\SansSerifBold10	at 5pt
=	\SansSerifItalics10	at 5pt
=	\SansSerifSlanted10	at 5pt
=				\Math10				at 5pt
=			\MathBold10			at 5pt
=			\Symbols10			at 5pt
=		\MoreSymbols10		at 5pt
=		\Typewriter10		at 5pt
=			\Gothic10			at 5pt
=			\Double10			at 5pt

= 			\Serif10 			at 7pt
= 		\SerifBold10 		at 7pt
= 	\SerifItalics10 	at 7pt
=	\SerifSlanted10 	at 7pt
=\SerifBoldItalics10	at 7pt
= 		\SansSerif10 		at 7pt
= 	\SansSerifBold10 	at 7pt
=\SansSerifItalics10	at 7pt
=\SansSerifSlanted10	at 7pt
=			\Math10				at 7pt
=		\MathBold10			at 7pt
=			\Symbols10			at 7pt
=		\MoreSymbols10		at 7pt
=		\Typewriter10		at 7pt
=			\Gothic10			at 7pt
=			\Double10			at 7pt

= 			\Serif10 			at 8pt
= 		\SerifBold10 		at 8pt
= 	\SerifItalics10 	at 8pt
=	\SerifSlanted10 	at 8pt
=\SerifBoldItalics10	at 8pt
= 		\SansSerif10 		at 8pt
= 	\SansSerifBold10 	at 8pt
=\SansSerifItalics10 at 8pt
=\SansSerifSlanted10 at 8pt
=			\Math10				at 8pt
=		\MathBold10			at 8pt
=			\Symbols10			at 8pt
=		\MoreSymbols10		at 8pt
=		\Typewriter10		at 8pt
=			\Gothic10			at 8pt
=			\Double10			at 8pt

= 			\Serif10 			at 10pt
= 		\SerifBold10 		at 10pt
= 		\SerifItalics10 	at 10pt
=		\SerifSlanted10 	at 10pt
=	\SerifBoldItalics10	at 10pt
= 		\SansSerif10 		at 10pt
= 	\SansSerifBold10 	at 10pt
= 	\SansSerifItalics10 at 10pt
= 	\SansSerifSlanted10 at 10pt
=				\Math10				at 10pt
=			\MathBold10			at 10pt
=			\Symbols10			at 10pt
=		\MoreSymbols10		at 10pt
=		\Typewriter10		at 10pt
=			\Gothic10			at 10pt
=			\Double10			at 10pt
=			\Relazioni10			at 10pt

= 				\Serif10 			at 12pt
= 			\SerifBold10 		at 12pt
= 		\SerifItalics10 	at 12pt
=		\SerifSlanted10 	at 12pt
=	\SerifBoldItalics10	at 12pt
= 			\SansSerif10 		at 12pt
= 		\SansSerifBold10 	at 12pt
= 	\SansSerifItalics10 at 12pt
= 	\SansSerifSlanted10 at 12pt
=				\Math10				at 12pt
=			\MathBold10			at 12pt
=			\Symbols10			at 12pt
=		\MoreSymbols10		at 12pt
=			\Typewriter10		at 12pt
=				\Gothic10			at 12pt
=				\Double10			at 12pt

= 			\Serif10 			at 14pt
= 		\SerifBold10 		at 14pt
= 	\SerifItalics10 	at 14pt
=		\SerifSlanted10 	at 14pt
=	\SerifBoldItalics10	at 14pt
= 		\SansSerif10 		at 14pt
= 	\SansSerifBold10 	at 14pt
= \SansSerifSlanted10 at 14pt
= \SansSerifItalics10 at 14pt
=				\Math10				at 14pt
=			\MathBold10			at 14pt
=			\Symbols10			at 14pt
=		\MoreSymbols10		at 14pt
=		\Typewriter10		at 14pt
=			\Gothic10			at 14pt
=			\Double10			at 14pt

\def\NormalStyle{\parindent=5pt\parskip=3pt\normalbaselineskip=14pt%
\def\nt{\tenSerif}%
\def\rm{\fam0\tenSerif}%
\textfont0=\tenSerif\scriptfont0=\sevenSerif\scriptscriptfont0=\fiveSerif
\textfont1=\tenMath\scriptfont1=\sevenMath\scriptscriptfont1=\fiveMath
\textfont2=\tenSymbols\scriptfont2=\sevenSymbols\scriptscriptfont2=\fiveSymbols
\textfont3=\tenMoreSymbols\scriptfont3=\sevenMoreSymbols\scriptscriptfont3=\fiveMoreSymbols
\textfont\itfam=\tenSerifItalics\def\it{\fam\itfam\tenSerifItalics}%
\textfont\slfam=\tenSerifSlanted\def\sl{\fam\slfam\tenSerifSlanted}%
\textfont\ttfam=\tenTypewriter\def\tt{\fam\ttfam\tenTypewriter}%
\textfont\bffam=\tenSerifBold%
\def\bf{\fam\bffam\tenSerifBold}\scriptfont\bffam=\sevenSerifBold\scriptscriptfont\bffam=\fiveSerifBold%
\def\cal{\tenSymbols}%
\def\greekbold{\tenMathBold}%
\def\gothic{\tenGothic}%
\def\Bbb{\tenDouble}%
\def\LieFont{\tenSerifItalics}%
\nt\normalbaselines\baselineskip=14pt%
}

\def\TitleStyle{\parindent=0pt\parskip=0pt\normalbaselineskip=15pt%
\def\nt{\fourteenSansSerifBold}%
\def\rm{\fam0\fourteenSansSerifBold}%
\textfont0=\fourteenSansSerifBold\scriptfont0=\tenSansSerifBold\scriptscriptfont0=\eightSansSerifBold
\textfont1=\fourteenMath\scriptfont1=\tenMath\scriptscriptfont1=\eightMath
\textfont2=\fourteenSymbols\scriptfont2=\tenSymbols\scriptscriptfont2=\eightSymbols
\textfont3=\fourteenMoreSymbols\scriptfont3=\tenMoreSymbols\scriptscriptfont3=\eightMoreSymbols
\textfont\itfam=\fourteenSansSerifItalics\def\it{\fam\itfam\fourteenSansSerifItalics}%
\textfont\slfam=\fourteenSansSerifSlanted\def\sl{\fam\slfam\fourteenSerifSansSlanted}%
\textfont\ttfam=\fourteenTypewriter\def\tt{\fam\ttfam\fourteenTypewriter}%
\textfont\bffam=\fourteenSansSerif%
\def\bf{\fam\bffam\fourteenSansSerif}\scriptfont\bffam=\tenSansSerif\scriptscriptfont\bffam=\eightSansSerif%
\def\cal{\fourteenSymbols}%
\def\greekbold{\fourteenMathBold}%
\def\gothic{\fourteenGothic}%
\def\Bbb{\fourteenDouble}%
\def\LieFont{\fourteenSerifItalics}%
\nt\normalbaselines\baselineskip=15pt%
}

\def\PartStyle{\parindent=0pt\parskip=0pt\normalbaselineskip=15pt%
\def\nt{\fourteenSansSerifBold}%
\def\rm{\fam0\fourteenSansSerifBold}%
\textfont0=\fourteenSansSerifBold\scriptfont0=\tenSansSerifBold\scriptscriptfont0=\eightSansSerifBold
\textfont1=\fourteenMath\scriptfont1=\tenMath\scriptscriptfont1=\eightMath
\textfont2=\fourteenSymbols\scriptfont2=\tenSymbols\scriptscriptfont2=\eightSymbols
\textfont3=\fourteenMoreSymbols\scriptfont3=\tenMoreSymbols\scriptscriptfont3=\eightMoreSymbols
\textfont\itfam=\fourteenSansSerifItalics\def\it{\fam\itfam\fourteenSansSerifItalics}%
\textfont\slfam=\fourteenSansSerifSlanted\def\sl{\fam\slfam\fourteenSerifSansSlanted}%
\textfont\ttfam=\fourteenTypewriter\def\tt{\fam\ttfam\fourteenTypewriter}%
\textfont\bffam=\fourteenSansSerif%
\def\bf{\fam\bffam\fourteenSansSerif}\scriptfont\bffam=\tenSansSerif\scriptscriptfont\bffam=\eightSansSerif%
\def\cal{\fourteenSymbols}%
\def\greekbold{\fourteenMathBold}%
\def\gothic{\fourteenGothic}%
\def\Bbb{\fourteenDouble}%
\def\LieFont{\fourteenSerifItalics}%
\nt\normalbaselines\baselineskip=15pt%
}

\def\ChapterStyle{\parindent=0pt\parskip=0pt\normalbaselineskip=15pt%
\def\nt{\fourteenSansSerifBold}%
\def\rm{\fam0\fourteenSansSerifBold}%
\textfont0=\fourteenSansSerifBold\scriptfont0=\tenSansSerifBold\scriptscriptfont0=\eightSansSerifBold
\textfont1=\fourteenMath\scriptfont1=\tenMath\scriptscriptfont1=\eightMath
\textfont2=\fourteenSymbols\scriptfont2=\tenSymbols\scriptscriptfont2=\eightSymbols
\textfont3=\fourteenMoreSymbols\scriptfont3=\tenMoreSymbols\scriptscriptfont3=\eightMoreSymbols
\textfont\itfam=\fourteenSansSerifItalics\def\it{\fam\itfam\fourteenSansSerifItalics}%
\textfont\slfam=\fourteenSansSerifSlanted\def\sl{\fam\slfam\fourteenSerifSansSlanted}%
\textfont\ttfam=\fourteenTypewriter\def\tt{\fam\ttfam\fourteenTypewriter}%
\textfont\bffam=\fourteenSansSerif%
\def\bf{\fam\bffam\fourteenSansSerif}\scriptfont\bffam=\tenSansSerif\scriptscriptfont\bffam=\eightSansSerif%
\def\cal{\fourteenSymbols}%
\def\greekbold{\fourteenMathBold}%
\def\gothic{\fourteenGothic}%
\def\Bbb{\fourteenDouble}%
\def\LieFont{\fourteenSerifItalics}%
\nt\normalbaselines\baselineskip=15pt%
}

\def\SectionStyle{\parindent=0pt\parskip=0pt\normalbaselineskip=13pt%
\def\nt{\twelveSansSerifBold}%
\def\rm{\fam0\twelveSansSerifBold}%
\textfont0=\twelveSansSerifBold\scriptfont0=\eightSansSerifBold\scriptscriptfont0=\eightSansSerifBold
\textfont1=\twelveMath\scriptfont1=\eightMath\scriptscriptfont1=\eightMath
\textfont2=\twelveSymbols\scriptfont2=\eightSymbols\scriptscriptfont2=\eightSymbols
\textfont3=\twelveMoreSymbols\scriptfont3=\eightMoreSymbols\scriptscriptfont3=\eightMoreSymbols
\textfont\itfam=\twelveSansSerifItalics\def\it{\fam\itfam\twelveSansSerifItalics}%
\textfont\slfam=\twelveSansSerifSlanted\def\sl{\fam\slfam\twelveSerifSansSlanted}%
\textfont\ttfam=\twelveTypewriter\def\tt{\fam\ttfam\twelveTypewriter}%
\textfont\bffam=\twelveSansSerif%
\def\bf{\fam\bffam\twelveSansSerif}\scriptfont\bffam=\eightSansSerif\scriptscriptfont\bffam=\eightSansSerif%
\def\cal{\twelveSymbols}%
\def\bg{\twelveMathBold}%
\def\gothic{\twelveGothic}%
\def\Bbb{\twelveDouble}%
\def\LieFont{\twelveSerifItalics}%
\nt\normalbaselines\baselineskip=13pt%
}

\def\SubSectionStyle{\parindent=0pt\parskip=0pt\normalbaselineskip=13pt%
\def\nt{\twelveSansSerifItalics}%
\def\rm{\fam0\twelveSansSerifItalics}%
\textfont0=\twelveSansSerifItalics\scriptfont0=\eightSansSerifItalics\scriptscriptfont0=\eightSansSerifItalics%
\textfont1=\twelveMath\scriptfont1=\eightMath\scriptscriptfont1=\eightMath%
\textfont2=\twelveSymbols\scriptfont2=\eightSymbols\scriptscriptfont2=\eightSymbols%
\textfont3=\twelveMoreSymbols\scriptfont3=\eightMoreSymbols\scriptscriptfont3=\eightMoreSymbols%
\textfont\itfam=\twelveSansSerif\def\it{\fam\itfam\twelveSansSerif}%
\textfont\slfam=\twelveSansSerifSlanted\def\sl{\fam\slfam\twelveSerifSansSlanted}%
\textfont\ttfam=\twelveTypewriter\def\tt{\fam\ttfam\twelveTypewriter}%
\textfont\bffam=\twelveSansSerifBold%
\def\bf{\fam\bffam\twelveSansSerifBold}\scriptfont\bffam=\eightSansSerifBold\scriptscriptfont\bffam=\eightSansSerifBold%
\def\cal{\twelveSymbols}%
\def\greekbold{\twelveMathBold}%
\def\gothic{\twelveGothic}%
\def\Bbb{\twelveDouble}%
\def\LieFont{\twelveSerifItalics}%
\nt\normalbaselines\baselineskip=13pt%
}

\def\AuthorStyle{\parindent=0pt\parskip=0pt\normalbaselineskip=14pt%
\def\nt{\tenSerif}%
\def\rm{\fam0\tenSerif}%
\textfont0=\tenSerif\scriptfont0=\sevenSerif\scriptscriptfont0=\fiveSerif
\textfont1=\tenMath\scriptfont1=\sevenMath\scriptscriptfont1=\fiveMath
\textfont2=\tenSymbols\scriptfont2=\sevenSymbols\scriptscriptfont2=\fiveSymbols
\textfont3=\tenMoreSymbols\scriptfont3=\sevenMoreSymbols\scriptscriptfont3=\fiveMoreSymbols
\textfont\itfam=\tenSerifItalics\def\it{\fam\itfam\tenSerifItalics}%
\textfont\slfam=\tenSerifSlanted\def\sl{\fam\slfam\tenSerifSlanted}%
\textfont\ttfam=\tenTypewriter\def\tt{\fam\ttfam\tenTypewriter}%
\textfont\bffam=\tenSerifBold%
\def\bf{\fam\bffam\tenSerifBold}\scriptfont\bffam=\sevenSerifBold\scriptscriptfont\bffam=\fiveSerifBold%
\def\cal{\tenSymbols}%
\def\greekbold{\tenMathBold}%
\def\gothic{\tenGothic}%
\def\Bbb{\tenDouble}%
\def\LieFont{\tenSerifItalics}%
\nt\normalbaselines\baselineskip=14pt%
}

\def\AddressStyle{\parindent=0pt\parskip=0pt\normalbaselineskip=14pt%
\def\nt{\eightSerif}%
\def\rm{\fam0\eightSerif}%
\textfont0=\eightSerif\scriptfont0=\sevenSerif\scriptscriptfont0=\fiveSerif
\textfont1=\eightMath\scriptfont1=\sevenMath\scriptscriptfont1=\fiveMath
\textfont2=\eightSymbols\scriptfont2=\sevenSymbols\scriptscriptfont2=\fiveSymbols
\textfont3=\eightMoreSymbols\scriptfont3=\sevenMoreSymbols\scriptscriptfont3=\fiveMoreSymbols
\textfont\itfam=\eightSerifItalics\def\it{\fam\itfam\eightSerifItalics}%
\textfont\slfam=\eightSerifSlanted\def\sl{\fam\slfam\eightSerifSlanted}%
\textfont\ttfam=\eightTypewriter\def\tt{\fam\ttfam\eightTypewriter}%
\textfont\bffam=\eightSerifBold%
\def\bf{\fam\bffam\eightSerifBold}\scriptfont\bffam=\sevenSerifBold\scriptscriptfont\bffam=\fiveSerifBold%
\def\cal{\eightSymbols}%
\def\greekbold{\eightMathBold}%
\def\gothic{\eightGothic}%
\def\Bbb{\eightDouble}%
\def\LieFont{\eightSerifItalics}%
\nt\normalbaselines\baselineskip=14pt%
}

\def\AbstractStyle{\parindent=0pt\parskip=0pt\normalbaselineskip=12pt%
\def\nt{\eightSerif}%
\def\rm{\fam0\eightSerif}%
\textfont0=\eightSerif\scriptfont0=\sevenSerif\scriptscriptfont0=\fiveSerif
\textfont1=\eightMath\scriptfont1=\sevenMath\scriptscriptfont1=\fiveMath
\textfont2=\eightSymbols\scriptfont2=\sevenSymbols\scriptscriptfont2=\fiveSymbols
\textfont3=\eightMoreSymbols\scriptfont3=\sevenMoreSymbols\scriptscriptfont3=\fiveMoreSymbols
\textfont\itfam=\eightSerifItalics\def\it{\fam\itfam\eightSerifItalics}%
\textfont\slfam=\eightSerifSlanted\def\sl{\fam\slfam\eightSerifSlanted}%
\textfont\ttfam=\eightTypewriter\def\tt{\fam\ttfam\eightTypewriter}%
\textfont\bffam=\eightSerifBold%
\def\bf{\fam\bffam\eightSerifBold}\scriptfont\bffam=\sevenSerifBold\scriptscriptfont\bffam=\fiveSerifBold%
\def\cal{\eightSymbols}%
\def\greekbold{\eightMathBold}%
\def\gothic{\eightGothic}%
\def\Bbb{\eightDouble}%
\def\LieFont{\eightSerifItalics}%
\nt\normalbaselines\baselineskip=12pt%
}

\def\RefsStyle{\parindent=0pt\parskip=0pt%
\def\nt{\eightSerif}%
\def\rm{\fam0\eightSerif}%
\textfont0=\eightSerif\scriptfont0=\sevenSerif\scriptscriptfont0=\fiveSerif
\textfont1=\eightMath\scriptfont1=\sevenMath\scriptscriptfont1=\fiveMath
\textfont2=\eightSymbols\scriptfont2=\sevenSymbols\scriptscriptfont2=\fiveSymbols
\textfont3=\eightMoreSymbols\scriptfont3=\sevenMoreSymbols\scriptscriptfont3=\fiveMoreSymbols
\textfont\itfam=\eightSerifItalics\def\it{\fam\itfam\eightSerifItalics}%
\textfont\slfam=\eightSerifSlanted\def\sl{\fam\slfam\eightSerifSlanted}%
\textfont\ttfam=\eightTypewriter\def\tt{\fam\ttfam\eightTypewriter}%
\textfont\bffam=\eightSerifBold%
\def\bf{\fam\bffam\eightSerifBold}\scriptfont\bffam=\sevenSerifBold\scriptscriptfont\bffam=\fiveSerifBold%
\def\cal{\eightSymbols}%
\def\greekbold{\eightMathBold}%
\def\gothic{\eightGothic}%
\def\Bbb{\eightDouble}%
\def\LieFont{\eightSerifItalics}%
\nt\normalbaselines\baselineskip=10pt%
}



%
%


\def\ModeYes{yes}
\def\ModeNo{no}

\def\ModeUndef{undefined}


\def\nx{\noexpand}
\def\ni{\noindent}
\def\newpage{\vfill\eject}

\def\ss{\vskip 5pt}
\def\ms{\vskip 10pt}
\def\bs{\vskip 20pt}

 \def\,{\mskip\thinmuskip}
 \def\!{\mskip-\thinmuskip}
 \def\>{\mskip\medmuskip}
 \def\;{\mskip\thickmuskip}

%
% ... 
%

\def\refsModePost{post}
\def\refsModeAuto{auto}

\def\dbRefsSatusModeOk{ok}
\def\dbRefsSatusModeError{error}
\def\dbRefsSatusModeWarning{warning}


\newcount\BNUM
\BNUM=0

\def\refs{}

\def\SetModePost{\xdef\refsMode{\refsModePost}}			
\SetModePost

\def\dbRefsStatusOk{%
	\xdef\dbRefsStatus{\dbRefsSatusModeOk}%
	\xdef\dbRefsError{\ModeNo}%
	\xdef\dbRefsWarning{\ModeNo}%
	\xdef\dbRefsInfo{\ModeNo}%
}

\def\dbRefs{%
}

\def\dbRefsGet#1{%
	\xdef\found{N}\xdef\ikey{#1}\dbRefsStatusOk%
	\xdef\key{\ModeUndef}\xdef\tag{\ModeUndef}\xdef\tail{\ModeUndef}%
	\dbRefs%
}

\def\NextRefsTag{%
	\global\advance\BNUM by 1%
}
\def\ShowTag#1{{\bf [#1]}}

\def\dbRefsInsert#1#2{%
\dbRefsGet{#1}%
\if\found Y %
   \xdef\dbRefsStatus{\dbRefsSatusModeWarning}%
   \xdef\dbRefsWarning{record is already there}%
   \xdef\dbRefsInfo{record not inserted}%
\else%
   \toks2=\expandafter{\dbRefs}%
   \ifx\refsMode\refsModeAuto \NextRefsTag
    \xdef\dbRefs{%
   	\the\toks2 \nx\xdef\nx\dbx{#1}%
	\nx\ifx\nx\ikey %
		\nx\dbx\nx\xdef\nx\found{Y}%
		\nx\xdef\nx\key{#1}%
		\nx\xdef\nx\tag{\the\BNUM}%
		\nx\xdef\nx\tail{#2}%
	\nx\fi}%
	\global\xdef\refs{\refs \ss\ni[\the\BNUM]\ #2\par}
   \fi%
   \ifx\refsMode\refsModePost 
    \xdef\dbRefs{%
   	\the\toks2 \nx\xdef\nx\dbx{#1}%
	\nx\ifx\nx\ikey %
		\nx\dbx\nx\xdef\nx\found{Y}%
		\nx\xdef\nx\key{#1}%
		\nx\xdef\nx\tag{\ModeUndef}%
		\nx\xdef\nx\tail{#2}%
	\nx\fi}%
   \fi%
\fi%
}

\def\dbRefsEdit#1#2#3{\dbRefsGet{#1}%
\if\found N 
   \xdef\dbRefsStatus{\dbRefsSatusModeError}%
   \xdef\dbRefsError{record is not there}%
   \xdef\dbRefsInfo{record not edited}%
\else%
   \toks2=\expandafter{\dbRefs}%
   \xdef\dbRefs{\the\toks2%
   \nx\xdef\nx\dbx{#1}%
   \nx\ifx\nx\ikey\nx\dbx %
	\nx\xdef\nx\found{Y}%
	\nx\xdef\nx\key{#1}%
	\nx\xdef\nx\tag{#2}%
	\nx\xdef\nx\tail{#3}%
   \nx\fi}%
\fi%
}

\def\bib#1#2{\RefsStyle\dbRefsInsert{#1}{#2}%
	\ifx\dbRefsStatus\dbRefsSatusModeWarning %
		\message{^^J}%
		\message{WARNING: Reference [#1] is doubled.^^J}%
	\fi%
}

\def\ref#1{\dbRefsGet{#1}%
\ifx\found N %
  \message{^^J}%
  \message{ERROR: Reference [#1] unknown.^^J}%
  \ShowTag{??}%
\else%
	\ifx\tag\ModeUndef \NextRefsTag%
		\dbRefsEdit{#1}{\the\BNUM}{\tail}%
		\dbRefsGet{#1}%
		\global\xdef\refs{\refs \ss\ni [\tag]\ \tail\par}
	\fi
	\ShowTag{\tag}%
\fi%
}

\def\ShowBiblio{\ms\Ensure{\SectionEnsure}%
{\SectionStyle\ni References}%
{\RefsStyle\refs}%
}

\newcount\CHANGES
\CHANGES=0
\def\AuxFile{7}
\def\PreventDoubleOn{\xdef\PreventDoubleLabel{\ModeYes}}

\PreventDoubleOn

\def\StoreLabel#1#2{\xdef\itag{#2}
 \ifx\PreModeStatus\ModeNo %
   \message{^^J}%
   \errmessage{You can't use Check without starting with OpenPreMode (and finishing with ClosePreMode)^^J}%
 \else%
   \immediate\write\AuxFile{\nx\dbLabelPreInsert{#1}{\itag}}%
   \dbLabelGet{#1}%
   \ifx\itag\tag %
   \else%
	\global\advance\CHANGES by 1%
 	\xdef\itag{(?.??)}%
    \fi%
   \fi%
}

\def\PreModeStatus{\ModeNo}

\def\ClosePreMode{\immediate\closeout\AuxFile%
  \ifnum\CHANGES=0%
	\message{^^J}%
	\message{**********************************^^J}%
	\message{**  NO CHANGES TO THE AuxFile  **^^J}%
	\message{**********************************^^J}%
 \else%
	\message{^^J}%
	\message{**************************************************^^J}%
	\message{**  PLAEASE TYPESET IT AGAIN (\the\CHANGES)  **^^J}%
    \errmessage{**************************************************^^ J}%
  \fi%
  \edef\PreModeStatus{\ModeNo}%
}

\def\dbLabelSatusModeOk{ok}

\def\dbLabelSatusModeWarning{warning}

\def\dbLabelStatusOk{%
	\xdef\dbLabelStatus{\dbLabelSatusModeOk}%
	\xdef\dbLabelError{\ModeNo}%
	\xdef\dbLabelWarning{\ModeNo}%
	\xdef\dbLabelInfo{\ModeNo}%
}

\def\dbLabel{%
}

\def\dbLabelGet#1{%
	\xdef\found{N}\xdef\ikey{#1}\dbLabelStatusOk%
	\xdef\key{\ModeUndef}\xdef\tag{\ModeUndef}\xdef\pre{\ModeUndef}%
	\dbLabel%
}

\def\ShowLabel#1{%
 \dbLabelGet{#1}%
 \ifx\tag \ModeUndef %
 	\global\advance\CHANGES by 1%
 	(?.??)%
 \else%
 	\tag%
 \fi%
}

\def\dbLabelPreInsert#1#2{\dbLabelGet{#1}%
\if\found Y %
  \xdef\dbLabelStatus{\dbLabelSatusModeWarning}%
   \xdef\dbLabelWarning{Label is already there}%
   \xdef\dbLabelInfo{Label not inserted}%
   \message{^^J}%
   \errmessage{Double pre definition of label [#1]^^J}%
\else%
   \toks2=\expandafter{\dbLabel}%
    \xdef\dbLabel{%
   	\the\toks2 \nx\xdef\nx\dbx{#1}%
	\nx\ifx\nx\ikey %
		\nx\dbx\nx\xdef\nx\found{Y}%
		\nx\xdef\nx\key{#1}%
		\nx\xdef\nx\tag{#2}%
		\nx\xdef\nx\pre{\ModeYes}%
	\nx\fi}%
\fi%
}

\def\dbLabelInsert#1#2{\dbLabelGet{#1}%
\xdef\itag{#2}%
\dbLabelGet{#1}%
\if\found Y %
	\ifx\tag\itag %
	\else%
	   \ifx\PreventDoubleLabel\ModeYes %
		\message{^^J}%
		\errmessage{Double definition of label [#1]^^J}%
	   \else%
		\message{^^J}%
		\message{Double definition of label [#1]^^J}%
	   \fi%
	\fi%
   \xdef\dbLabelStatus{\dbLabelSatusModeWarning}%
   \xdef\dbLabelWarning{Label is already there}%
   \xdef\dbLabelInfo{Label not inserted}%
\else%
   \toks2=\expandafter{\dbLabel}%
    \xdef\dbLabel{%
   	\the\toks2 \nx\xdef\nx\dbx{#1}%
	\nx\ifx\nx\ikey %
		\nx\dbx\nx\xdef\nx\found{Y}%
		\nx\xdef\nx\key{#1}%
		\nx\xdef\nx\tag{#2}%
		\nx\xdef\nx\pre{\ModeNo}%
	\nx\fi}%
\fi%
}


\newcount\PART
\newcount\CHAPTER
\newcount\SECTION
\newcount\SUBSECTION
\newcount\FNUMBER

\PART=0
\CHAPTER=0
\SECTION=0
\SUBSECTION=0	
\FNUMBER=0

\def\LastPart{\ModeUndef}
\def\LastChapter{\ModeUndef}
\def\LastSection{\ModeUndef}
\def\LastSubSection{\ModeUndef}
\def\LastClaim{\ModeUndef}
\def\Last{\ModeUndef}

\newdimen\TOBOTTOM
\newdimen\LIMIT

\def\Ensure#1{\ \par\ \immediate\LIMIT=#1\immediate\TOBOTTOM=\the\pagegoal\advance\TOBOTTOM by -\pagetotal%
\ifdim\TOBOTTOM<\LIMIT\newpage \else%
\vskip-\parskip\vskip-\parskip\vskip-\baselineskip\fi}

\def\PartLabel{\the\PART}
\def\NewPart#1{\global\advance\PART by 1%
         \bs\ni{\PartStyle  Part \PartLabel:}
         \bs\ni{\PartStyle #1}\newpage%
         \CHAPTER=0\SECTION=0\SUBSECTION=0\FNUMBER=0%
         \gdef\Left{#1}%
         \global\edef\Last{\PartLabel}%
         \global\edef\LastPart{\PartLabel}%
         \global\edef\LastChapter{\ModeUndef}%
         \global\edef\LastSection{\ModeUndef}%
         \global\edef\LastSubSection{\ModeUndef}%
         \global\edef\LastClaim{\ModeUndef}}
\def\ChapterLabel{\the\CHAPTER}
\def\NewChapter#1{\global\advance\CHAPTER by 1%
         \bs\ni{\ChapterStyle  Chapter \ChapterLabel: #1}\ms%
         \SECTION=0\SUBSECTION=0\FNUMBER=0%
         \gdef\Left{#1}%
         \global\edef\Last{\ChapterLabel}%
         \global\edef\LastChapter{\ChapterLabel}%
         \global\edef\LastSection{\ModeUndef}%
         \global\edef\LastSubSection{\ModeUndef}%
         \global\edef\LastClaim{\ModeUndef}}
\def\SectionEnsure{3cm}
\def\NewSection#1{\Ensure{\SectionEnsure}\gdef\SectionLabel{\the\SECTION}\global\advance\SECTION by 1%
         \ms\ni{\SectionStyle  \SectionLabel.\ #1}\ss%
         \SUBSECTION=0\FNUMBER=0%
         \gdef\Left{#1}%
         \global\edef\Last{\SectionLabel}%
         \global\edef\LastSection{\SectionLabel}%
         \global\edef\LastSubSection{\ModeUndef}%
         \global\edef\LastClaim{\ModeUndef}}
\def\NewAppendix#1#2{\Ensure{\SectionEnsure}\gdef\SectionLabel{#1}\global\advance\SECTION by 1%
         \bs\ni{\SectionStyle  Appendix \SectionLabel.\ #2}\ss%
         \SUBSECTION=0\FNUMBER=0%
         \gdef\Left{#2}%
         \global\edef\Last{\SectionLabel}%
         \global\edef\LastSection{\SectionLabel}%
         \global\edef\LastSubSection{\ModeUndef}%
         \global\edef\LastClaim{\ModeUndef}}
\def\Acknowledgements{\Ensure{\SectionEnsure}\gdef\SectionLabel{}%
         \ms\ni{\SectionStyle  Acknowledgments}\ss%
         \SECTION=0\SUBSECTION=0\FNUMBER=0%
         \gdef\Left{}%
         \global\edef\Last{\ModeUndef}%
         \global\edef\LastSection{\ModeUndef}%
         \global\edef\LastSubSection{\ModeUndef}%
         \global\edef\LastClaim{\ModeUndef}}
\def\SubSectionEnsure{2cm}
\def\SubSectionLabel{\ifnum\SECTION>0 \the\SECTION.\fi\the\SUBSECTION}
\def\NewSubSection#1{\Ensure{\SubSectionEnsure}\global\advance\SUBSECTION by 1%
         \ms\ni{\SubSectionStyle #1}\ss%
         \global\edef\Last{\SubSectionLabel}%
         \global\edef\LastSubSection{\SubSectionLabel}}
\def\SetNumberingModeN{\def\ClaimLabel{(\the\FNUMBER)}}
\def\SetNumberingModeSN{\def\ClaimLabel{(\ifnum\SECTION>0 \SectionLabel.\fi%
      \the\FNUMBER)}}
\def\SetNumberingModeCSN{\def\ClaimLabel{(\ifnum\CHAPTER>0 \the\CHAPTER.\fi%
      \ifnum\SECTION>0 \SectionLabel.\fi%
      \the\FNUMBER)}}

\def\NewClaim{\global\advance\FNUMBER by 1%
    \ClaimLabel%
    \global\edef\LastClaim{\ClaimLabel}%
    \global\edef\Last{\ClaimLabel}}

\def\HideLabels{\xdef\ShowLabelsMode{\ModeNo}}
\HideLabels

\def\fn{\eqno{\NewClaim}} 
\def\fl#1{%
\ifx\ShowLabelsMode\ModeYes%
 \eqno{{\buildrel{\hbox{\AbstractStyle[#1]}}\over{\hfill\NewClaim}}}%
\else%
 \eqno{\NewClaim}%
\fi%
\dbLabelInsert{#1}{\ClaimLabel}}
\def\fprel#1{\global\advance\FNUMBER by 1\StoreLabel{#1}{\ClaimLabel}%
\ifx\ShowLabelsMode\ModeYes%
\eqno{{\buildrel{\hbox{\AbstractStyle[#1]}}\over{\hfill.\itag}}}%
\else%
 \eqno{\itag}%
\fi%
}

\def\cl#1{\global\advance\FNUMBER by 1\dbLabelInsert{#1}{\ClaimLabel}%
\ifx\ShowLabelsMode\ModeYes%
${\buildrel{\hbox{\AbstractStyle[#1]}}\over{\hfill\ClaimLabel}}$%
\else%
  $\ClaimLabel$%
\fi%
}
\def\cprel#1{\global\advance\FNUMBER by 1\StoreLabel{#1}{\ClaimLabel}%
\ifx\ShowLabelsMode\ModeYes%
${\buildrel{\hbox{\AbstractStyle[#1]}}\over{\hfill.\itag}}$%
\else%
  $\itag$%
\fi%
}

\def\Note{\ms\leftskip 3cm\rightskip 1.5cm\AbstractStyle}
\def\endNote{\par\leftskip 2cm\rightskip 0cm\NormalStyle\ss}


\parindent=7pt
\leftskip=2cm
\newcount\SideIndent
\newcount\SideIndentTemp
\SideIndent=0
\newdimen\SectionIndent
\SectionIndent=-8pt

\def\sidebar{\vrule height15pt width.2pt }
\def\endcorner{\hbox{\hbox{\vrule height6pt width.2pt}\vbox to6pt{\vfill\hbox
to4pt{\leaders\hrule height0.2pt\hfill}}}}
\def\begincorner{\hbox{\hbox{\vrule height6pt width.2pt}\vbox to6pt{\hbox
to4pt{\leaders\hrule height0.2pt\hfill}}}}
\def\endbegincorner{\hbox{\vbox to15pt{\endcorner\vskip-6pt\begincorner\vfill}}}
\def\SideShow{\SideIndentTemp=\SideIndent \ifnum \SideIndentTemp>0 
\loop\sidebar\hskip 2pt \advance\SideIndentTemp by-1\ifnum \SideIndentTemp>1 \repeat\fi}

\def\BeginSection{{\vbadness 100000 \par\ni\hskip\SectionIndent%
\SideShow\vbox to 15pt{\vfill\begincorner}}\global\advance\SideIndent by1\vskip-10pt}

\def\EndSection{{\vbadness 100000 \par\ni\global\advance\SideIndent by-1%
\hskip\SectionIndent\SideShow\vbox to15pt{\endcorner\vfill}\vskip-10pt}}

\def\EndBeginSection{{\vbadness 100000\par\ni%
\global\advance\SideIndent by-1\hskip\SectionIndent\SideShow
\vbox to15pt{\vfill\endbegincorner}}%
\global\advance\SideIndent by1\vskip-10pt}

\def\ShowBeginCorners#1{%
\SideIndentTemp =#1 \advance\SideIndentTemp by-1%
\ifnum \SideIndentTemp>0 %
\vskip-15truept\hbox{\kern 2truept\vbox{\hbox{\begincorner}%
\ShowBeginCorners{\SideIndentTemp}\vskip-3truept}}%
\fi%
}

\def\ShowEndCorners#1{%
\SideIndentTemp =#1 \advance\SideIndentTemp by-1%
\ifnum \SideIndentTemp>0 %
\vskip-15truept\hbox{\kern 2truept\vbox{\hbox{\endcorner}%
\ShowEndCorners{\SideIndentTemp}\vskip 2truept}}%
\fi%
}

\def\BeginSections#1{{\vbadness 100000 \par\ni\hskip\SectionIndent%
\SideShow\vbox to 15pt{\vfill\ShowBeginCorners{#1}}}\global\advance\SideIndent by#1\vskip-10pt}

\def\EndSections#1{{\vbadness 100000 \par\ni\global\advance\SideIndent by-#1%
\hskip\SectionIndent\SideShow\vbox to15pt{\vskip15pt\ShowEndCorners{#1}\vfill}\vskip-10pt}}

\def\EndBeginSections#1#2{{\vbadness 100000\par\ni%
\global\advance\SideIndent by-#1%
\hbox{\hskip\SectionIndent\SideShow\kern-2pt%
\vbox to15pt{\vskip15pt\ShowEndCorners{#1}\vskip4pt\ShowBeginCorners{#2}}}}%
\global\advance\SideIndent by#2\vskip-10pt}




%
%


\def\al{\alpha}
\def\be{\beta}
\def\de{\delta}

\def\ep{\epsilon}

\def\om{\omega}
\def\si{\sigma}

\def\ka{\kappa}

\def\La{\Lambda}


 \def\su{{\hbox{\gothic su}}}

 \def\spin{{\hbox{\gothic spin}}}


 \def\L{{\hbox{\Bbb L}}}


\def\det{{\hbox{det}}}

\def\Diff{{\hbox{Diff}}}

\def\ip{\hbox to4pt{\leaders\hrule height0.3pt\hfill}\vbox to8pt{\leaders\vrule width0.3pt\vfill}\kern 2pt}
 
\def\del{\partial}
\def\na{\nabla}

\def\arr{\rightarrow}

\def\then{\Rightarrow}

%
%

\def\cases#1{\left\{\eqalign{#1}\right.}
\NormalStyle
\SetNumberingModeSN
\PreventDoubleOn

\long\def\title#1{\centerline{\TitleStyle\ni#1}}

\long\def\author#1{\ms\centerline{\AuthorStyle by {\it #1}}}

\long\def\address#1{\ss\centerline{\AddressStyle #1}\par}
\long\def\moreaddress#1{\centerline{\AddressStyle #1}\par}
\def\abstract{\ms\leftskip 3cm\rightskip .5cm\AbstractStyle{\bf \ni Abstract:}\ }
\def\endabstract{\par\leftskip 2cm\rightskip 0cm\NormalStyle\ss}

\SetNumberingModeSN

\def\nab#1{{\buildrel #1\over \na}}
\def\frac[#1/#2]{\hbox{$#1\over#2$}}
\def\Frac[#1/#2]{{#1\over#2}}
\def\({\left(}
\def\){\right)}
\def\[{\left[}
\def\]{\right]}
\def\^#1{{}^{#1}_{\>\cdot}}
\def\_#1{{}_{#1}^{\>\cdot}}
\def\Label=#1{{\buildrel {\hbox{\fiveSerif \ShowLabel{#1}}}\over =}}
\def\<{\kern -1pt}


\def\ExpandAllCNotes{\long\def\CNote##1{%
\BeginSection
	\Note%
 		##1%
	\endNote%
\EndSection%
}}
\ExpandAllCNotes
%
%
%
%


\def\frame#1{\vbox{\hrule\hbox{\vrule\vbox{\kern2pt\hbox{\kern2pt#1\kern2pt}\kern2pt}\vrule}\hrule\kern-4pt}}

\def\Box to #1#2#3{\frame{\vtop{\hbox to #1{\hfill #2 \hfill}\hbox to #1{\hfill #3 \hfill}}}}


\bib{Nester}{C.\ M.\ Chen, J.\ M.\ Nester,  
{\it Quasilocal quantities for GR and other gravity theories},
Class.Quant.Grav. 16 (1999) 1279-1304;\goodbreak
C.-M. Chen, J.M. Nester,
Gravitation \& Cosmology {\bf 6}, (2000), 257  (gr--qc/0001088)
}

\bib{Augmented}{L.~Fatibene, M.~Ferraris, M.~Francaviglia,
{\it Augmented Variational Principles and Relative Conservation Laws in Classical Field Theory},
{  Int. J. Geom. Methods Mod. Phys.}, {\bf 2}(3), (2005), pp. 373-392; [math-ph/0411029v1]
}

\bib{Sinicco}{M.\ Ferraris, M.\ Francaviglia and I.\ Sinicco, Il Nuovo Cimento, {\bf 107B},(11), 1992, 1303
}

\bib{ADM}{R.  Arnowitt, S.  Deser and C.  W.  Misner, in: 
{\it Gravitation: An Introduction to Current Research}, 
L. Witten ed. Wyley,  227, (New York, 1962)}

\bib{RT}{T.\ Regge, C.\ Teitelboim, Annals of Physics {\bf 88}, 286  (1974).
}

\bib{Wald}{L.ÊFatibene, M. Ferraris, M.Francaviglia, M. Raiteri, 
{\it Remarks on N\"other charges and black holes entropy.}  
Ann. Physics 275(1) (1999)
}

\bib{Kerr2}{{L. Fatibene, M. Ferraris, M. Francaviglia, S.Mercadante,
{\it About the Energy of AdS-Kerr Solutions},
Int. J. Geom. Methods Mod. Phys. 5(7), 2008.}
}

\bib{Kerr}{L. Fatibene, M.Ferraris, M. Francaviglia,
{\it The Energy of a Solution from Different Lagrangians},
{ Int. J. Geom. Methods Mod. Phys.}, {\bf 3}(7), (2006), pp. 1341-1347
}

\bib{Book}{L. Fatibene, M. Francaviglia,
{\it Natural and Gauge Natural Formalism for Classical Field Theories},
Kluwer Academic Publishers, (Dordrecht, 2003), xxii
}

\bib{OurKosman}{L. Fatibene, M. Ferraris, M. Francaviglia, M. Godina,
{\it A geometric definition of Lie derivative for Spinor Fields},
in: Proceedings of 
{\it ``6th International Conference on Differential Geometry
and its Applications, August 28--September 1, 1995"}, (Brno, Czech Republic),
Editor: I. Kol{\'a}{\v r}, MU University, Brno, Czech Republic (1996)}

\bib{Kosman}{Kosmann, Y., (1972), Ann. di Matematica Pura et Appl. {\bf 91} 317--395.\goodbreak
Kosmann, Y., (1966), Comptes Rendus Acad. Sc. Paris, s\'erie A, {\bf 262} 289--292.\goodbreak
Kosmann, Y., (1966), Comptes Rendus Acad. Sc. Paris, s\'erie A, {\bf 262} 394--397.\goodbreak
Kosmann, Y., (1967), Comptes Rendus Acad. Sc. Paris, s\'erie A, {\bf 264} 355--358.
}

\bib{Cavalese}{M.~Ferraris, M.~Francaviglia, 
in: {\sl 8th Italian Conference on General Relativity and    Gravitational Physics}, 
Cavalese (Trento), August 30--September 3, World Scientific, Singapore, 1988, 183}

\bib{Maple}{K.Chu, C.Farel, G.Fee, R.McLenaghan, Fields Inst. Comm. {\bf 15}, (1997)
}

\bib{Julia}{B. Julia and S. Silva, 
{\it Currents and superpotentials in classical gauge theories}, 
Classical Quantum Gravity , vol. 17, No. 22 (2000), 4733-4743}

\bib{Torre}{I.M. Anderson, C.G. Torre, Phys. Rev. Lett. 77 (1996) 4109 (hepÐ th/9608008); \goodbreak
C.G. Torre, hepÐth/9706092, Lectures given at 2nd Mexican School on Gravitation and Mathematical Physics, Tlaxcala, Mexico (1996)}

\bib{BY}{L.\ Fatibene, M.\ Ferraris, M.\ Francaviglia, M.\ Raiteri, J.  Math. Phys., {\bf 42}, No. 3, 1173 (2001)  (gr--qc/0003019). }

\bib{BTZ}{L.~Fatibene, M.~Ferraris, M.~Francaviglia, M.~Raiteri, 
{\sl Remarks on conserved      quantities and entropy of BTZ black hole solutions.  I. The general setting.} 
Phys.\ Rev.\ D (3) 60 (1999), n.12, 124012
}

\bib{Taub}{L. Fatibene, M. Ferraris, M. Francaviglia, M. Raiteri,
{\it The entropy of the Taub-bolt solution},
Ann. Physics 284 (2000), no. 2, 197Ð214.}

\bib{Taub2}{R. Clarkson, L. Fatibene, R.B. Mann,
{\it Thermodynamics of $(d+1)$-dimensional NUT-charged AdS spacetimes},
Nuclear Phys. B 652 (2003), no. 1-3, 348Ð382.}

\bib{Weinberg}{{S. Weinberg,
{\it Gravitation and Cosmology: Principles and Applications of the General Theory of Relativity}}
Wiley, New York (a.o.) (1972). XXVIII, 657 S. : graph. Darst.. ISBN: 0-471-92567-5.}

\bib{BL}{Joseph Katz, Jiri Bicak, and Donald Lynden-Bell.
{\it Relativistic conservation laws and integral constraints for large  cosmological perturbations},
{Phys. Rev. D}, 55:5957--5969, 1997.}

\bib{Landau}{L. Landau, E. Lifchitz,
{\it Th\'eorie du champ},
Moscou, ƒditions Mir. 1966.
}

\bib{Petrov}{A.N.Petrov,
{\it Nonlinear Perturbations and Conservation Laws on Curved Backgrounds in GR and Other Metric Theories},
; arXiv:0705.0019v1
}

\bib{Freud}{{\bf Freud}
{\it }
}

\def\AppA{A}

\def\ubal{\underline{\al}\kern1pt}
\def\obal{\overline{\al}\kern1pt}

\def\ubR{\underline{R}\kern1pt}
\def\obR{\overline{R}\kern1pt}
\def\ubom{\underline{\om}\kern1pt}
\def\obxi{\overline{\xi}\kern1pt}
\def\ubu{\underline{u}\kern1pt}
\def\ube{\underline{e}\kern1pt}
\def\obe{\overline{e}\kern1pt}

\def\beR{{}^\be\<\<R}
\def\sgn{\hbox{sgn}}

\def\AppA{A}

\NormalStyle

\title{Extended Loop Quantum Gravity\footnote{$^*$}{{\AbstractStyle
	This paper is published despite the effects of the Italian law 133/08 ({\tt http://groups.google.it/group/scienceaction}). 
        This law drastically reduces public funds to public Italian universities, which is particularly dangerous for free scientific research, 
        and it will prevent young researchers from getting a position, either temporary or tenured, in Italy.
        The authors are protesting against this law to obtain its cancellation.\goodbreak}}}

\author{L.Fatibene$^{a, b}$, M.Ferraris$^a$, M.Francaviglia$^{a,b, c}$}

\address{$^a$ Department of Mathematics, University of Torino (Italy)}

\moreaddress{$^b$ INFN - Iniziativa Specifica Na12}

\moreaddress{$^c$ LCS, University of Calabria (Italy)}

\bib{Barbero}{F.\ Barbero, 
{\it Real Ashtekar variables for Lorentzian signature space-time},
Phys.\ Rev.\ {\it D51}, 5507, 1996}

\bib{Immirzi}{G.\ Immirzi, 
{\it Quantum Gravity and Regge Calculus},
Nucl.\ Phys.\ Proc.\ Suppl.\ {\bf 57}, 65-72}

\bib{Rovelli}{C.\ Rovelli,
{\it Quantum Gravity}, 
Cambridge University Press, Cambridge, 2004}

\bib{Capozziello}{S. Capozziello, V.F. Cardone, V. Salzano,
{\it Cosmography of $f(R)$ gravity},
Phys.Rev.D{\bf 78}, 063504, 2008
}

\bib{Thieman}{T. Thiemann,
{\it LoopQuantumGravity: An InsideView}, 
hep-th/0608210}

\bib{UniversalityBV}{A. Borowiec, M. Ferraris, M. Francaviglia, I. Volovich,
{\it Universality of Einstein Equations for the Ricci Squared Lagrangians},
Class. Quantum Grav. 15, 43-55, 1998}

\bib{Odintzov}{S. Nojiri, S.D. Odintsov,
{\it Modified gravity as realistic candidate for dark energy, inflation and dark matter}
AIP Conf.Proc. 1115, 2009, 212-217;
arXiv:0810.1557}

\bib{Magnano}{G. Magnano, M. Ferraris, M. Francaviglia, 
{\it Nonlinear gravitational Lagrangians},
Gen.Rel.Grav. {\bf 19}(5), 1987, 465-479}

\bib{Bojowald}{M. Bojowald,
{\it Consistent Loop Quantum Cosmology}
Class.Quant.Grav.{\bf 26} 075020, 2009}

\bib{Libro}{L.\ Fatibene, M.\ Francaviglia, 
{\it Natural and gauge natural formalism for classical field theories. A geometric perspective including spinors and gauge theories}, 
Kluwer Academic Publishers, Dordrecht, 2003}

\bib{Spin2005}{L.\ Fatibene, M.\ Francaviglia, 
{\it Spin Structures on Manifolds and Ashtekar Variables}, 
Int. J. Geom. Methods Mod. Phys. {\bf 2}(2), 147-157, (2005)}

\bib{Universality}{L.Fatibene, M.Ferraris, M.Francaviglia,
{\it New Cases of Universality Theorem for Gravitational Theories},
(in preparation)}

\NormalStyle

\abstract
We discuss constraint structure of extended theories of gravitation (also known as $f(R)$ theories) in the vacuum selfdual formulation
introduced in \ref{Universality}.
\endabstract

\NewSection{Introduction}

We have recently investigated a formulation of $f(R)$ theories (in a metric-affine framework) based on non-linear actions 
similar to the Holst Lagrangian; see \ref{Universality}.
These actions are in fact written in terms of the scalar curvature $\beR$ of the Barbero-Immirzi connection with parameter $\be$ (see \ref{Barbero}, \ref{Immirzi})
and are dynamically equivalent to the corresponding ``classical'' $f(R)$ theory. For the linear case $f(\beR)=\beR$ one obtains the standard Holst action.
Hence these new actions are to be understood as Barbero-Immirzi formulations of the corresponding classical $f(R)$ theory.

This could be interesting for at least two reasons: from the point of view of LQG this new formulation provides a family of models which are classically well--understood and investigated in detail (see \ref{Capozziello}, \ref{Odintzov}). There are many classical effects known in $f(R)$ theories that should be traced in their quantum genesis.
The minisuperspace of these models is quite well--understood and should be studied in loop quantum cosmology (LQC) formulation (see \ref{Bojowald}),
to contribute to a better understanding of the classical limit of LQG models.
Moreover, as in all metric-affine models, matter has a non-trivial feedback on the gravitational field which would be also interesting to trace in its quantum
origin. It is often said that matter in LQG {\it simply} adds new labels to spin networks, while in these models one could expect a more complicated mechanism that would be certainly interesting to be discussed in detail.
Finally, there are a number of equivalences, e.g.~with scalar tensor models (see \ref{Magnano}), that again would be interesting to be discussed in detail at quantum level. Let us stress that these equivalences are known to hold at the classical level and, as usual, one should investigate whether they still hold at the full quantum level or just emerge classically.

From the classical viewpoint we shall here provide a route to define a quantization {\it \`a la} loop of $f(R)$ theories. 
Of course classical effects of these extended theories of gravitation have been extensively investigated.
It  is therefore interesting to investigate also their quantum effects. For example it would be interesting to see whether the removal of singularities that has been shown to hold in standard loop quantization of GR is preserved generically in these extended gravitational models.

For the sake of simplicity we shall here restrict our attention to the Euclidean signature and to the selfdual formulation (which in the Euclidean sector is in fact a special case of the Barbero-Immirzi formulation) and show that one can apply LQG methods (see \ref{Rovelli}) also to the quantization of these theories.
In vacuum we shall obtain something similar to Einstein gravity with a cosmological constant. 
This is very well expected on the basis of a classical equivalence (see \ref{UniversalityBV}); however, let us stress that our result seems to establish a stronger equivalence at the {\it quantum} level and not only at the classical level.

Moreover, let us stress that the classical equivalence holds only in vacuum, while  the equivalence is broken when generic matter is considered and the extended models are equivalent to scalar tensor theories; see \ref{Magnano}. Tracing the mechanism which leads to this shift of equivalence at the quantum level would be  therefore rather interesting and will be investigated in forthcoming papers.
We shall follow the notation introduced in \ref{Universality} and \ref{Rovelli}.

\NewSection{Selfdual Formulation for Extended Theories}

In \ref{Universality} we introduced 
$$
{}^\be\<R:= R^{ab}{}_{\mu\nu}  e_a^\mu e_b^\nu   +\be R^{ab}{}_{\mu\nu}   e^{c\mu} e^{d\nu} \ep_{cdab}
\fn$$
where $e_a^\mu$ is a spin frame (see \ref{Libro}), $R^{ab}{}_{\mu\nu}$ is the curvature of a spin connection $\om^{ab}_\mu$ on a $4$ dimensional (spin) manifold $M$ and $\be\not=0$ is a real parameter. Indices $a, c, \dots= 0..3$ and $\mu,\nu,\dots= 0..3$ while $i,j, \dots=1..3$. 

In the Euclidean sector  one obtains for $\be=\frac[1/2]$ the standard selfdual curvature 
$$
{}^+\<R:=R^{ab}{}_{\mu\nu}  e_a^\mu e_b^\nu   +\frac[1/2] R^{ab}{}_{\mu\nu}   e^{c\mu} e^{d\nu} \ep_{cdab}
\fn$$
which can be written in terms of the curvature $F^i_{\mu\nu}:=p^i_{ab} R^{ab}{}_{\mu\nu}$ of the usual selfdual connection $A^i_\mu:= p^i_{ab} \om^{ab}_\mu=\om^{0i}_\mu + \frac[1/2] \ep^i{}_{jk} \om^{jk}_\mu$ as follows
$$
\frac[1/2]{}^+\<R=  \frac[1/2]  R^{cd}{}_{\mu\nu}\( \de^a_{[c} \de^b_{d]}    + \frac[1/2] \ep_{cd}{}^{ab}\)e_a^{\mu} e_b^{\nu}=
R^{cd}{}_{\mu\nu} p^i_{cd} p_i^{ab} e_a^{\mu} e_b^{\nu}
 = p_i^{ab} F^i_{\mu\nu} e^\mu_a e^\nu_b =: F
\fn$$

\CNote{
Here $p_i^{ab}$ denotes the algebra projector $p:\spin(4)\arr \su(2)$ on selfdual forms.
It is given by
$$
p_i^{0j}= \frac[1/2] \de^j_i
\qquad
p_i^{j0}= -p_i^{0j}
\qquad
p_i^{jk}= \frac[1/2] \ep_i{}^{jk}
\fn$$
and the inverse projector $p^i_{ab}$ is defined by
$$
p^i_{0j}= \frac[1/2] \de_j^i
\qquad
p^i_{j0}= -p^i_{0j}
\qquad
p^i_{jk}= \frac[1/2] \ep^i{}_{jk}
\fn$$
One can easily prove that 
$$
p^i_{ab} p^{ab}_j= \de^i_j
\qquad
p^i_{ab} p^{cd}_i= \frac[1/2]\( \de^a_{[c}\de^b_{d]} +\frac[1/2] \ep^{ab}{}_{cd}\)
\fn$$
}

One is then led to consider the following family of Lagrangians
$$
L^+= \frac[1/2\ka] e f(F) + L_m
\fl{LagF}$$
where $\ka=8\pi G$,  $e$ is the determinant of the frame matrix, $f$ is a generic analytic function and $L_m$ encodes the matter contribution.
Usually matter is assumed to couple only with $g$ (and possibly to its derivatives up to some finite order; usually, in view of minimal coupling principle, at most $1$) and not to the connection $\om^{ab}_\mu$. 
Hereafter we shall just consider the vacuum sector, i.e.~we set $L_m=0$.

In the special case $f(F)=F$ one obtains an equivalent formulation of the usual selfdual action
$$
\eqalign{
\L^+=&\frac[1/8\ka] {}^+\<R^{ab}\land e^c\land e^d \ep_{abcd} =
\frac[1/16\ka]\(R^{ab}{}_{\mu\nu}  + \frac[1/2]\ep^{ab}{}_{ef} R^{ef}{}_{\mu\nu}\) e^c_\rho e^d_\si \ep^{\mu\nu\rho\si} \ep_{abcd} ds = \cr
=&
\frac[1/14\ka]R^{ef}{}_{\mu\nu} \( \de^a_{[e}\de^b_{f]}  + \frac[1/2]\ep^{ab}{}_{ef}\)e^c_\rho e^d_\si \ep^{\mu\nu\rho\si}  \ep_{abcd} ds=
\frac[1/8\ka]R^{ef}{}_{\mu\nu} p^{ab}_i p^i_{ef} e^c_\rho e^d_\si \ep^{\mu\nu\rho\si}  \ep_{abcd} ds=\cr
=& \frac[1/8\ka]p^{ab}_i  F^i{}_{\mu\nu}  e^c_\rho e^d_\si \ep^{\mu\nu\rho\si}  \ep_{abcd} ds=
\frac[e/2\ka]p^{ab}_i  F^i{}_{\mu\nu} e_a^\mu e_b^\nu ds=
\frac[e/2\ka]  F ds
}
\fn$$
where $ds$ is the standard local basis of $4$-forms on $M$ induced by coordinates.

Field equations of the Lagrangian $\L^+$ are
$$
\cases{
& p^{ab}_i F^i_{\mu\nu}  e^c_\rho \ep^{\mu\nu\rho\si} \ep_{abcd}=0\cr
& p^i_{ab}\na_{\mu}\( e^a_{\nu} e^b_{\rho}\)\ep^{\mu\nu\rho\si}=0\cr
}
\fl{FE}$$

Let us now consider a Cauchy (boundary) surface $i:S\arr M: k^A\mapsto x^\mu(k)$, $A,B, \dots=1..3$; 
in coordinates $x^\mu=(t, k^A)$ adapted to the submanifold $S$ one has $i:k^A\mapsto k^A$ and $\del_A x^\mu= \de^\mu_A$.

\CNote{The unit covector normal to $S$ is given by $n=dx^0$. One can use antiselfdual transformations to define a canonical adapted frame
$e_a=\ube_a^\mu \del_\mu$ and coframe $e^a=\obe^a_\mu d x^\mu$
(see \ref{Spin2005}) given by
$$
\cases{
&\ube_0^0=N^{-1}
\qquad\quad \ube_i^0=0\cr
&\ube_0^j=N^{-1} N^j
\quad\>\>\>\ube_i^j= \underline{\al}^i_j\cr
}
\qquad\qquad
\cases{
&\obe_0^0= N
\qquad\quad\quad\>\obe_i^0=0\cr
&\obe_0^j= -N^l \overline{\al}^j_l
\quad\quad\obe_i^j= \overline{\al}^i_j\cr
}
\fn$$
Tetrads (or better spin frames; \ref{Libro}) adapted to $S$ define triads $\ep_i=e_i= \underline{\al}_i^A \del_A$ on $S$. Also the selfdual connection can be projected onto $S$ to define a connection $A^i_A= A^i_\mu \del_A x^\mu$ on $S$.
Let us denote by $F^i_{AB}= F^i_{\mu\nu} \del_A x^\mu \del_B x^\nu$ the projected curvature (which is the same as the curvature of the projected connection);
for later convenience let us also define the tangent-normal projection of the curvature
$F^i_A= F^i_{\mu\nu} \del_A x^\mu n^\nu$ (of course the normal-normal projection vanishes due to the skew symmetry of $F$).
 
Let us also set $E^A_i= \ep \ep^A_i$ for the momentum conjugated to the connection $A^i_A$ written in terms of the triad $\ep_i^A$ tangent to $S$,
with $\ep$ the determinant of the (co)triad $\ep^i_A$.
}

Field equations \ShowLabel{FE} can be projected onto $S$ to obtain
a number of evolution equations and the following constraints on $S$:
$$
\cases{
& \nab{A}_A E^A_i=0\cr
& F^i_{AB} E^A_i=0\cr
&\ep_i{}^{jk} F^i_{AB} E^A_jE^B_k=0\cr
}
\fl{PreQuantumEqs}$$

These constitute the starting point of LQG quantization scheme; the first equation is related to gauge covariance, the second to 
$\Diff(S)$--covariance; while the third equation is called the {\it Hamiltonian constraint}, 
when quantized it becomes the so-called {\it Wheeler-deWitt equation} and it encodes the (quantum) dynamics.
In order to solve the first and second equation one defines an Hilbert space spanned by spin knots (see \ref{Rovelli}) so that the Wheeler-deWitt  equation is implemented as an operator on that space and it defines physical states.

On this basis one expects to be able to perform the same steps with extended models $f(F)$; since the extended models are still gauge and generally covariant,
the first and second equations are expected to remain unchanged. 
This would mean that the definition of Area and Volume operators are unchanged and ``spacetime'' gets discretized in extended models {\it exactly} as in standard LQG.
Since extended models are known to provide a modified dynamics with respect to standard GR
one also expect that the Wheeler-deWitt equation has to be modified.

We shall hereafter compute the analogous of equations \ShowLabel{PreQuantumEqs} for the action \ShowLabel{LagF} in order to fully confirm our expectations.

\NewSection{Constraint Structure}

Let us then consider the Lagrangian
$$
L^+= \frac[e/2\ka]  f(F)
\fn$$
i.e.~the purely gravitational part of \ShowLabel{LagF}.

Field equations are
$$
\cases{
& f' p^{ab}_i F^i_{\mu\nu} e_a^\mu -\frac[1/2] f e^b_\nu=0\cr
&p_i^{ab}\na_\mu\(e f' e_a^\mu e_b^\nu\)=0\cr
}
\fn$$
The {\it master equation} $f' F -2 f =0$ is obtained by tracing the first one by means of  $e_b^\nu$; see \ref{Universality} and \ref{UniversalityBV}.
This can be replaced back into the first equation to obtain
$$
 f' \(p^{ab}_i F^i_{\mu\nu} e_a^\mu -\frac[1/4]  F e^b_\nu\)=0
 \quad\then
 p^{ab}_i F^i_{\mu\nu} e_a^\mu -\frac[1/4]  F e^b_\nu=0
\fn$$
where we used the fact that generically $f'\not=0$ on the zeroes of the master equation.
For simplicity let us assume that the master equation has only one (simple) zero $F=\rho$; 
when there are many (simple) zeroes each of them defines a sector of the quantum theory and one is supposed to sum over all sectors,
which are in correspondence with the discrete zero structure of the analytic function $f$.

Let us also define a conformal tetrad $\tilde e^a_\mu= \sqrt{|f'|}e^a_\mu$, set $\si=\sgn(f'(\rho))$ and use tilde to denote quantities depending on
the conformal tetrad, e.g.~$\tilde E_i^A = \tilde \ep \tilde \ep_i^A= |f'| E_i^A$ and 
$$
\tilde F= p_i^{ab} F^i_{\mu\nu} \tilde e^\mu_a\tilde e^\nu_b= \frac[\si/f']F
\fn$$
 
Field equations are hence equivalent to
$$
\cases{
& p^{ab}_i F^i_{\mu\nu} \tilde e_a^\mu -\frac[1/4] \tilde F \tilde e^b_\nu=0\cr
&f' F -2 f =0
\qquad\then F=\rho\cr
&p_i^{ab}\na_\mu\(\tilde e  \tilde e_a^\mu \tilde e_b^\nu\)=0\cr
}
\fn$$
The third equation implies the constraint
$$
\nab{A}_A \tilde E^A_i=0
\fn$$
as in the standard case, though for the conformal frame $\tilde e^a_\mu$.

The second equation can be now expanded as
$$
\tilde F= p^{ab}_i F^i_{\mu\nu} \tilde e_a^\mu \tilde e_b^\nu=  2p^{0l}_i F^i_{\mu\nu} \tilde e_0^\mu \tilde e_l^\nu + p^{lk}_i F^i_{\mu\nu} \tilde e_l^\mu \tilde e_k^\nu=
 -\tilde F^i_{A}  \tilde \ep_i^A + \frac[1/2]\ep_i{}^{lk} F^i_{AB} \tilde\ep_l^A \tilde\ep_k^B= \frac[\si/f']\rho
\fl{F}$$
which allows us to express $\tilde F^i_{A} \tilde \ep_i^A$ as a function of constrained fields, i.e.
$$
\tilde F^i_{A}  \tilde\ep_i^A = \frac[1/2]\ep_i{}^{lk} F^i_{AB} \tilde\ep_l^A \tilde\ep_k^B -\frac[\si/f']\rho
\fl{Ftn}$$

Notice that  the first equation is really different from the standard case (i.e.~LQG without cosmological constant) 
due to the different coefficient $\frac[1/4]$ (which in the standard case is $\frac[1/2]$ and allows a complete cancellation of $\tilde F^i_{A}  \tilde\ep_i^A $). 
The standard case in LQG can be recovered by setting $f(F)=F$; in this case the master equation simply 
implies $F=0$ and the standard case without cosmological constant is obtained in particular.
The first equation can be projected in the normal direction to the constraint to obtain
$$
\(p^{ab}_i F^i_{\mu\nu} \tilde e_a^\mu  -\frac[1/4] \tilde F \tilde e^b_\nu \)\tilde e_b^\al \tilde e^\nu_d n_\al=0 \qquad\then
\fn$$
$$
p^{j0}_i F^i_{\mu\nu} \tilde e_j^\mu\tilde e^\nu_d   -\frac[1/4] \tilde F    \de^0_d=0 \qquad\then
\fn$$
$$
F^i_{A\nu} \tilde e_i^A \tilde e^\nu_d   +\frac[1/2] \tilde F    \de^0_d=0
\fn$$

For $d=k=1..3$ one has
$$
F^i_{AB} \tilde e_i^A \tilde e^B_k  =0 \qquad\then
F^i_{AB} \tilde E_i^A  =0 
\fn$$

For $d=0$ one has instead
$$
\tilde F^i_{A} \tilde e_i^A    +\frac[1/2] \tilde F   =0
\fn$$
and, using \ShowLabel{F} and \ShowLabel{Ftn}, one obtains
$$
\eqalign{
&\tilde F^i_{A} \tilde e_i^A    -\frac[1/2] \tilde F^i_{A}  \tilde \ep_i^A + \frac[1/4]\ep_i{}^{lk} F^i_{AB} \tilde \ep_l^A \tilde \ep_k^B  =
\frac[1/2] \tilde F^i_{A}  \tilde\ep_i^A + \frac[1/4]\ep_i{}^{lk} F^i_{AB} \tilde \ep_l^A \tilde\ep_k^B=\cr
&= \frac[1/4]\ep_i{}^{lk} F^i_{AB} \tilde\ep_l^A \tilde\ep_k^B - \frac[\si/2f']  \rho+ \frac[1/4]\ep_i{}^{lk} F^i_{AB} \tilde \ep_l^A \tilde\ep_k^B
= \frac[1/2]\ep_i{}^{lk} F^i_{AB} \tilde\ep_l^A \tilde\ep_k^B - \frac[\si/2f' ]  \rho=0
}
\fn$$

$$
\ep_i{}^{lk} F^i_{AB} \tilde\ep_l^A \tilde\ep_k^B= \frac[\si/f'] \rho
\fn$$

$$
\ep_i{}^{lk} F^i_{AB} \tilde E_l^A \tilde E_k^B= \frac[\si/f'] \rho \tilde \ep^2= \frac[\si/f']\rho\tilde E
\fn$$
where $\tilde E:= \det (\tilde \ep \tilde \ep_i^A) =\tilde \ep^{3} \tilde \ep^{-1}= \tilde \ep^2$ denotes the determinant of the conformal momentum  $\tilde E_i^A$.

Let us stress that all this can be done also in the standard LQG framework, though in that case $F^i_{A} $ does not enter other constraints 
and hence can be ignored.

Accordingly,  the constraints can be written in terms of the conformal triad as follows
$$
\cases{
& \nab{A}_A \tilde E^A_i=0\cr
& F^i_{AB} \tilde E^A_i=0\cr
&\ep_i{}^{jk} F^i_{AB} \tilde E^A_j \tilde E^B_k= \frac[\si/f'] \rho \tilde E\cr
}
\fl{Constrf}$$

As expected, the first and second constraints are unchanged with respect to \ShowLabel{PreQuantumEqs}, while the Wheeler-deWitt equation is modified 
by the ``source term''  $\frac[\si/f'] \rho \tilde E$, which explicitly  depends on the non-linearity of $f(F)$.
This is the quantum counterpart of what happens classically for $f(R)$ theories and reflects also what happens in standard LQG with the cosmological constant $\La=-\frac[1/4|f'|]\rho$; see Appendix \AppA.
Let us also notice that the third constraint is a density, which is fundamental in the approach to quantization proposed by Thiemann; see \ref{Thieman}.

\NewSection{Conclusions and Perspectives}

We have shown that, in the generic extended models introduced in \ref{Universality}, constraints allows a loop theory approach to quantization formally similar to
what one usually does in vacuum models with cosmological constant.
This shows that the equivalence between $f(R)$ models and Einstein with cosmological constant (shown in \ref{UniversalityBV} to hold in the classical theory)
holds also at the quantum level.

Of course more attention should be paid when matter couplings  are considered,  when this equivalence is known to break and is replaced at least by a conformal equivalence.

Also the whole Hamiltonian structure of the theory should be verified in detail to exclude second class constraints which might add further equations to
the set \ShowLabel{Constrf}. These constraints \ShowLabel{Constrf} are in any case necessary conditions on the boundary $S$.
Since from them discretization of ``spacetime'' follows one can claim in any event that extended spacetimes are discretized as in standard LQG.

\NewAppendix{\AppA}{LQG with Cosmological Constant}
 
 Let us here briefly review the standard result for LQG quantization in vacuum with cosmological constant in order to compare it with what we found for extended models.
 
 Let us consider the Lagrangian
 $$
 \eqalign{
 L_\La=& \({}^+\<\<R^{ab} +\frac[\La/6]e^a\land e^b\)\land e^c\land e^d \ep_{abcd}=
\( \frac[1/2] {}^+\<\<R^{ab}{}_{\mu\nu} +\frac[\La/6]e^a_\rho e^b_\si\) e^c_\rho  e^d_\si\ep^{\mu\nu\rho\si} \ep_{abcd} ds=\cr
=& e\(\frac[1/2] {}^+\<\<R^{ab}{}_{\mu\nu} e_e^\mu e_f^\nu \ep^{efcd} \ep_{abcd} +\frac[\La/6] \ep^{abcd}\ep_{abcd}\)ds
= 2e \({}^+\<\<R^{ab}{}_{\mu\nu} e_a^\mu e_b^\nu  +2\La  \) ds\cr
 }
 \fn$$
which can also be written in terms of the selfdual curvature as
$$
 L_\La= \( 2p^{ab}_i F^i +\frac[\La/6]e^a\land e^b\)\land e^c\land e^d \ep_{abcd }
 \fn$$

By varying this Lagrangian one gets the following field equations
$$
\cases{
&\(p^{ab}_i F^i_{\mu\nu} + \frac[4\La/6] e^a_\mu e^b_\nu\)e^c_\rho \ep^{\mu\nu\rho\si} \ep_{abcd}=0\cr
&p_i^{ab} \na_\mu \(e^c_\rho e^d_\si\)\ep^{\mu\nu\rho\si} \ep_{abcd}=0\cr
}
\fn$$ 

By projecting on the boundary $S$ one gets the following constraints
$$
\cases{
&\nab{A}_A E_i^A=0\cr
& F^i_{AB} E_i^A=0\cr
&\ep_i{}^{jk} F^i_{AB} E_i^A E_j^B = -4\La E\cr
}
\fn$$
which account for the value of the cosmological constant as claimed after \ShowLabel{Constrf}
in which, however, the conformal frame was used.

\Acknowledgements

We wish to thank C.~Rovelli for discussions about Barbero-Immirzi formulation.
We acknowledge the contribution of INFN (Iniziativa Specifica NA12) and 
the local research project {\it Leggi di conservazione in teorie della gravitazione classiche e quantistiche} (2010) 
of Dipartimento di Matematica of University of Torino (Italy).

\ShowBiblio

\end